\documentclass[prd,amsmath,amssymb,showpacs,eqsecnum]{revtex4}
\usepackage{latexsym}
\usepackage[dvips]{graphics}

\newcommand{\II}{\openone}

\newcommand{\kb}{{\boldsymbol{k}}}
\newcommand{\kpb}{{\boldsymbol{k'}}}
\newcommand{\sigb}{{\boldsymbol{\sigma}}}
\newcommand{\ZZ}{{\mathbb{Z}}}

\begin{document}

\title{Quantum Weak Energy Inequalities for the Dirac field in Flat Spacetime}
\author{C.J. Fewster}
\email{cjf3@york.ac.uk}
\author{B. Mistry}
\affiliation{Department of Mathematics, University of York, Heslington, York, YO10 5DD, UK}
\date{15 September 2003}
\begin{abstract}  
Quantum Weak Energy Inequalities (QWEIs) have been established for a
variety of quantum field theories in both flat and curved spacetimes.
Dirac fields are known (by a result of Fewster and Verch) to satisfy
QWEIs under very general circumstances. However this result does not
provide an explicit formula for the QWEI bound, so its magnitude has not
previously been determined. In this paper we present a new and explicit QWEI bound
for Dirac fields of mass $m\geqslant 0$ in four-dimensional Minkowski space.  
We follow the methods employed by Fewster and Eveson for the
scalar field, modified to take account of anticommutation
relations. A key ingredient is an identity for Fourier transforms
established by Fewster and Verch. We also compare our QWEI with those
previously obtained for scalar and spin-1 fields.
\end{abstract}
\pacs{03.70.+k, 11.10.Ef}
\maketitle

\section{Introduction}
It has been known for a long time that the energy density of a quantum
field may assume negative values~\cite{EGJ}. If the magnitude and the
duration of such negative energy densities were unconstrained various
exotic phenomema might be possible, ranging from the manufacture of
macroscopic traversable wormholes~\cite{FR-worm} to large-scale violations of the
second law of thermodynamics~\cite{Ford78}.

However, quantum field theory itself contains mechanisms to limit the
extent of negative energy densities: certain weighted averages of the
stress-energy tensor obey state-independent lower bounds known as
quantum weak energy inequalities (QWEI) [or simply quantum inequalities (QI)]
which severely limit exotic phenomena such as those described above.

Bounds of this type have been obtained by various 
means~\cite{fordroman3,FordRoman97,flan,PfenningFord98,FewsterEveson, 
CJFTeo,AGWQI,Pfenn,FewsterPfenning,Helfer2,Flan2} for the scalar,
electromagnetic and Proca fields, in both flat and curved spacetimes, 
leading to results of great generality. Taking four-dimensional
Minkowski space as a specific example, and
averaging the energy density along an inertial worldline, the QWEI
satisfied by these three theories may be written in the 
form~\cite{FewsterEveson,CJFTeo,Pfenn,FewsterPfenning}
\begin{equation}
\int dt\, {\langle :T_{00}: \rangle}_{\psi} (t, \boldsymbol{x_0})   {g(t)}^{2} 
\geqslant - \frac{\mathcal{S} }{16{\pi}^{3} }\  \int_{m}^{\infty} du \, |\widehat{g} (u) |^{2} u^{4} 
Q_{3}(u/m)
\label{eq:scalarQI}
\end{equation}
for any smooth, real-valued $g$ vanishing outside a compact region,
where $m$ is the particle's mass, $\mathcal{S}$ denotes its helicity ($\mathcal{S} =1$ for
scalars, $2$ for photons and $3$ for massive spin-$1$ particles), while
\begin{equation} 
Q_{3}(x) = \left( 1 - \frac{1}{x^2} \right)^{1/2} \left( 1 -
\frac{1}{2x^2}  \right)  -  \frac{1}{2x^4} \ln (x + \sqrt{x^{2} -
1})
\label{eq:Q3}
\end{equation}
and is replaced by $\lim_{x \rightarrow \infty} Q_{3}(x) = 1$ in the massless
case. (In curved spacetimes the QWEIs of these theories are not simply
related by an overall factor.)

In the spin-$\frac{1}{2}$ case only
two arguments are known.  One, due to Vollick~\cite{Vollick2} (based on earlier work
of Flanagan~\cite{flan}), exploits the conformal properties of massless
Dirac fields in two-dimensional spacetimes to obtain a
sharp (and explicit) QWEI for an arbitrary sampling
function.  The other, due to Fewster and Verch~\cite{CJFVerch},
establishes the existence of QEWIs for Dirac and Majorana fields of
arbitrary mass in general four-dimensional globally hyperbolic
spacetimes by drawing on techniques from microlocal analysis.  
However, this approach did not lead to an explicit formula for the QWEI bound, so its magnitude---even in 
four-dimensional Minkowski space---has not previously
been determined~\footnote{In
principle, one could obtain an explicit bound from~\cite{CJFVerch} by estimating
various constant quantities arising in the derivation.  However, this
would require considerable labour and is unlikely to provide a
particularly enlightening result.}. (Specific classes of Dirac states with local negative energy
densities have been studied~\cite{Vollick1,YuShu} and shown, in the
massless limit, to satisfy the scalar field QWEI obtained
in~\cite{FordRoman97}; of course, this does not amount to
a proof of the QWEIs.) Thus, for example, it has not previously been clear whether or
not it is more or less easy to support a traversable wormhole using 
Dirac fields, rather than with the scalar or electromagnetic fields.

In this paper, we present a new QWEI derivation for Dirac fields, which
combines ideas taken from~\cite{CJFVerch} with an approach first
developed for the scalar field in~\cite{FewsterEveson,CJFTeo}. The argument
is straightforward, involving the use of anticommutation relations and
an identity for Fourier transforms, and yields an explicit, closed-form, bound. In four-dimensional
Minkowski space this bound is
\begin{equation}
\int dt\, {\langle :T_{00}: \rangle}_{\psi} (t, \boldsymbol{x_0})  {g(t)}^{2} 
\geqslant - \frac{1}{12{\pi}^{3} }\  \int_{m}^{\infty} \ du \ |\widehat{g} (u) |^{2} \ u^{4} \
Q_{3}^{D} (u/m),
\end{equation}
where
\begin{equation}
Q_{3}^{D}(x) = 4   \left( 1 - \frac{1}{x^2} \right)^{3/2} - 3 Q_{3}(x)
\end{equation}
is replaced by $\lim_{x \rightarrow \infty} Q_{3}(x) = 1$ in the massless
case.
Although we have restricted ourselves, for simplicity, to
Minkowski space, our technique
certainly extends to static spacetimes and may even suggest a new
approach in the general globally hyperbolic case.  
It is hoped to return to these issues elsewhere.\\[0.2cm]

\noindent \textit{Conventions:}
We take the metric to have signature $({+}{-}{-}{-})$ and adopt units in which $\hbar = c = 1$.
The Fourier transform $\widehat{f}$ of a function $f$
on $\mathbb{R}$ is defined by 
\begin{equation}
\widehat{f}(\omega) = \int_{-\infty}^{\infty} dt \, f(t) e^{-i \omega t}.
\end{equation}

\section{The QWEI derivation}

To establish our notation, we begin by recalling some basic facts about the Dirac
fields. 
The Dirac equation for a fermion field $\psi$ of mass $m$ is
\begin{equation}
( i \gamma^{\mu} \partial_{\mu} - m ) \psi = 0, 
\end{equation}
where the $\gamma$-matrices are given in terms of the Pauli
matrices $\sigma_{i}$ by
\begin{equation}
\gamma^{0}=
\begin{pmatrix}
1 & 0 \\
0 & -1
\end{pmatrix}, \qquad
\gamma^{i}=
\begin{pmatrix}
0 & \sigma_{i}\\
-\sigma_{i} & 0   
\end{pmatrix}   \quad (i=1,2,3)
\end{equation}
and obey $\{ \gamma^{\mu}, \gamma^{\nu}  \} = 2 \eta^{\mu\nu}$. 
The stress-energy tensor~\cite{Belinfante} is
\begin{equation} 
T_{\mu \nu} = \frac{i}{4}[ \overline{\psi} \gamma_{\mu} \overleftrightarrow{\partial_{\nu}} \psi +
\overline{\psi} \gamma_{\nu}
\overleftrightarrow{\partial_{\mu}}
\psi ],
\end{equation}
where $\overline{\psi}=\psi^\dagger\gamma^0$, 
so the classical Dirac energy density is
\begin{equation}
T_{00} =\frac{i}{2}[\psi^{\dagger} \dot{\psi} - \dot{\psi^{\dagger}}\psi].
\label{eq:T00}
\end{equation}
(Of course, this quantity is unbounded both from above and below, in
contrast to the energy density of a classical scalar field. This is at
the root of the differences between the QWEIs obeyed by the scalar and Dirac
fields.)
To simplify our notation, we quantise the Dirac field in a box of side $L$
and then take the continuum limit $L\to \infty$ towards the end of the
argument. We emphasise that this is done purely for notational
convenience. The Dirac field operator $\psi(x)$ is therefore 
defined in terms of creation and
annihilation operators by
\begin{equation} 
\psi(x) = \sum_{\kb}\ \sum_{\alpha=1,2} \ 
[ b_{\alpha}(\kb) u^{\alpha}(\kb) e^{-ik\cdot x} +
d^{\dagger}_{\alpha} (\kb)  v^{\alpha} (\kb) e^{ik\cdot x}], 
\label{eq:psi}  
\end{equation}
where $(L/\pi)\kb$ runs through $\ZZ^3$, and the four-vector $k$ has components
$k^a=(\omega_\kb,\kb)$ with $\omega_\kb =\sqrt{\|\kb\|^{2}+m^{2}}$. The 
annihilation and creation operators satisfy the
anticommutation relations
\begin{equation}
\{ b_\alpha(\kb), b_{\alpha'}^{\dagger}(\kpb) \}
     =  \delta_{\alpha,\alpha'} \delta_{\kb,\kpb} \II
\label{eq:ACR1}
\end{equation}
and
\begin{equation}
\{ d_\alpha(\kb), d_{\alpha'}^{\dagger}(\kpb) \} =
      \delta_{\alpha,\alpha'} \delta_{\kb,\kpb} \II,
\label{eq:ACR2} 
\end{equation} 
with all other
anticommutators vanishing. Finally, 
the spinors $u^{\alpha}(\kb)$ and $v^{\alpha}(\kb)$, where
$\alpha=1,2$ labels the two independent spin states, are given by
\begin{equation}
u^{\alpha}(\kb)=
\begin{pmatrix}
\displaystyle \sqrt{\frac{\omega_\kb+m}{2 \omega_\kb V}} \phi^{\alpha}    \\[0.4cm]
\displaystyle    \frac{\sigb \cdot \kb}{\sqrt{2 \omega_\kb (\omega_\kb + m) V}} \phi^{\alpha}
\end{pmatrix} \quad\hbox{and}\quad
 v^{\alpha}(\kb)=
\begin{pmatrix}
\displaystyle  \frac{\sigb \cdot \kb}{\sqrt{2 \omega_\kb (\omega_\kb + m) V}} \phi^{\alpha}     \\[0.4cm]
\displaystyle \sqrt{\frac{\omega_\kb+m}{2 \omega_\kb V}} \phi^{\alpha} 
\end{pmatrix},
\end{equation}
in which the two dimensional column vectors $\phi^{\alpha}$ are
$\phi^{1\dagger}=(1,0)$ and  $\phi^{2\dagger}=(0,1)$ and $V=L^3$. The normalisation
has been chosen so that
\begin{equation}
\sum_\alpha \|u^\alpha(\kb)\|^2 = \sum_\alpha \|v^\alpha(\kb)\|^2
=\frac{2}{V}\,.
\label{eq:normalisation}
\end{equation}

Substituting the quantum field $\psi(x)$ into the classical energy
density Eq.~\eqref{eq:T00} and normal ordering according to the
prescription
\begin{equation}
{:}d_{\alpha}(\kb)d^{\dagger}_{\alpha'}(\kpb){:} = -d^{\dagger}_{\alpha'}(\kpb)d_{\alpha}(\kb),
\end{equation}
we obtain the normal ordered energy density at the spatial
origin $(t,\boldsymbol{0})$ in the form      
\begin{eqnarray}
{:}T_{00}{:}(t,\boldsymbol{0}) &=& \frac{1}{2}   \sum_{\kb,\kpb}  \sum_{\alpha,\alpha'}
 \{  (\omega_{\kb}+\omega_{\kpb})   
[ b_\alpha^\dagger (\kb) b_{\alpha'}(\kpb)  u^{\alpha \dagger}(\kb)   
u^{\alpha'}(\kpb)e^{i(\omega_{\kb}-\omega_{\kpb})t} \nonumber\\
&&\indent + 
 d_{\alpha'}^{\dagger} (\kpb)  d_\alpha (\kb)    v^{\alpha \dagger}(\kb)   
v^{\alpha'}(\kpb)e^{-i(\omega_{\kb}-\omega_{\kpb})t}] \nonumber \\  
&&\indent  + (\omega_{\kpb}-\omega_{\kb})  [   d_{\alpha} (\kb)  b_{\alpha'} (\kpb)    v^{\alpha \dagger}(\kb)   
u^{\alpha'}(\kpb)e^{-i(\omega_{\kb}+\omega_{\kpb})t} \nonumber \\  
&&\indent- b_\alpha^\dagger (\kb) d_{\alpha'}^{\dagger}(\kpb)   u^{\alpha \dagger}(\kb)   
v^{\alpha'}(\kpb)e^{i(\omega_{\kb}+\omega_{\kpb})t}]  \}.   
\end{eqnarray} 
We will consider weighted energy density averages of the form 
\begin{equation}
T_{f}=
\int_{-\infty}^{\infty} dt \, {:}T_{00}{:}  (t, \boldsymbol{0}) f(t),
\end{equation}
measured by a stationary observer at spacetime origin $\boldsymbol{0}$,
where $f$ is a non-negative sampling
function. Thus, 
\begin{eqnarray}
  T_{f}  &=& \frac{1}{2}   \sum_{\kb,\kpb}  \sum_{\alpha,\alpha'}
 \{  (\omega_{\kb}+\omega_{\kpb})   
[ b_\alpha^\dagger (\kb) b_{\alpha'}(\kpb)   u^{\alpha \dagger}(\kb)   
u^{\alpha'}(\kpb) \widehat{f}(\omega_{\kpb}-\omega_{\kb})  \nonumber\\
&&\indent+ 
 d_{\alpha'}^{\dagger} (\kpb)  d_\alpha (\kb)    v^{\alpha \dagger}(\kb)   
v^{\alpha'}(\kpb)   \widehat{f}(\omega_{\kb}-\omega_{\kpb}) ] \nonumber \\  
&&\indent  + (\omega_{\kpb}-\omega_{\kb})  [   d_{\alpha} (\kb)  b_{\alpha'} (\kpb)    v^{\alpha \dagger}(\kb)   
u^{\alpha'}(\kpb)  \widehat{f}(\omega_{\kb}+\omega_{\kpb}) \nonumber\\
&&\indent-
 b_\alpha^\dagger (\kb) d_{\alpha'}^{\dagger}(\kpb)   u^{\alpha \dagger}(\kb)   
v^{\alpha'}(\kpb)  \widehat{f} (-\omega_{\kb}-\omega_{\kpb})]  \}. 
\label{eq:Tf}   
\end{eqnarray} 
Our aim is to determine a lower bound on expectation values of
$T_{f}$ in the case where $f=g^2$, for some real-valued, smooth, compactly
supported function $g$. To this end, we first define a family $\{\mathcal{O}_{\mu i} | \mu \in
\mathbb{R}, i=1,2,3,4 \}$ of
Fock space operators by
\begin{equation}
\mathcal{O}_{\mu i}= \sum_{\alpha',\kpb} \{ \overline{\widehat{g}(-\omega_{\kpb} +
\mu)}  b_{\alpha'}(\kpb) u^{\alpha'}_i(\kpb) +
\overline{\widehat{g}(\omega_{\kpb}+\mu)} d_{\alpha'}^{\dagger}(\kpb)
v^{\alpha'}_i(\kpb) \}.
\end{equation}
A short calculation, using the anticommutation relations and the
normalisation~\eqref{eq:normalisation}, shows that
\begin{eqnarray}
 \mathcal{O}^{\dagger}_{\mu i}\mathcal{O}_{\mu i} &=& S_\mu \II +
 \sum_{\alpha,\alpha'} \sum_{\kb,\kpb} \left\{  \widehat{g}(-\omega_{\kb}+\mu)
\overline{ \widehat{g}(-\omega_{\kpb} + \mu) } b^{\dagger}_\alpha(\kb)
b_{\alpha'}(\kpb) u^{\alpha \dagger}(\kb) u^{\alpha'}(\kpb)   \right.  \nonumber\\ 
&&\phantom{S_\mu \II  \sum\sum}
- \widehat{g}(\omega_{\kb}+\mu)
\overline{ \widehat{g}(\omega_{\kpb} + \mu) }
d^{\dagger}_{\alpha'}(\kpb) d_\alpha(\kb) v^{\alpha \dagger}(\kb)
v^{\alpha'}(\kpb)\nonumber\\
&&\phantom{S_\mu \II  \sum\sum} 
+ \widehat{g}(\omega_{\kb}+\mu)
\overline{ \widehat{g}(-\omega_{\kpb} + \mu) } d_\alpha(\kb)
b_{\alpha'}(\kpb) v^{\alpha \dagger}(\kb) u^{\alpha'}(\kpb)  \nonumber\\
&&\phantom{S_\mu \II \sum\sum}
\left.+ \widehat{g}(-\omega_{\kb}+\mu)
\overline{ \widehat{g}(\omega_{\kpb} + \mu) } b^{\dagger}_\alpha(\kb)
d^{\dagger}_{\alpha'}(\kpb) u^{\alpha \dagger}(\kb) v^{\alpha'}(\kpb)  
\right\}
\end{eqnarray} 
where 
\begin{equation}
S_\mu = \frac{2}{V} \sum_{\kb}
|\widehat{g}(\omega_{\kb}+\mu)|^{2}, 
\label{eq:Smu}           
\end{equation}
and we have implicitly summed over the spinor index $i$. 

To relate this to our expression for $T_{f}$ we use the
following Lemma, which plays an analogous role here to that played by
the convolution theorem in~\cite{FewsterEveson,CJFTeo}.\\

{\noindent\bf Lemma 1:} {\em
If $f=g^2$ for some real-valued, smooth, compactly supported $g\in C_0^{\infty}( \mathbb{R})$
then}
\begin{equation}
 (\omega+\omega')\widehat{f}(\omega-\omega') =  
\int_{-\infty}^\infty  \frac{d\mu}{\pi}\, \mu \widehat{g}(\omega-\mu)
\overline{\widehat{g}(\omega'-\mu)}.   
\end{equation} 
This Lemma was also used in the QWEI derivation in~\cite{CJFVerch}; 
for completeness we give a proof of this statement in the
Appendix which differs slightly from that given in that reference.  
Using this Lemma with the anticommutation relations Eq.~\eqref{eq:ACR1}
and Eq.~\eqref{eq:ACR2}, we calculate
\begin{eqnarray}
\int_{-\infty}^\infty d\mu\, \mu (\mathcal{O}^{\dagger}_\mu\mathcal{O}_\mu-S_\mu\II)
&=& 
\pi \sum_{\alpha,\alpha'} \sum_{\kb,\kpb}  \{  (\omega_{\kb}+\omega_{\kpb})[ 
b^{\dagger}_{\alpha} (\kb)
b_{\alpha'}(\kpb)u^{\alpha \dagger}(\kb)u^{\alpha'}(\kpb)
\widehat{f}(\omega_{\kpb}-\omega_{\kb}) \notag \\
&&\phantom{\sum\sum} +
d^{\dagger}_{\alpha'} (\kpb)
d_{\alpha}(\kb)v^{\alpha \dagger}(\kb)v^{\alpha'}(\kpb)
\widehat{f}(\omega_{\kb}-\omega_{\kpb})] \notag \\
&&\phantom{\sum\sum} +  
(\omega_{\kpb}-\omega_{\kb})[ d_{\alpha} (\kb)
b_{\alpha'}(\kpb)v^{\alpha \dagger}(\kb)u^{\alpha'}(\kpb)
\widehat{f}(\omega_{\kb}+\omega_{\kpb}) \notag \\
&&\phantom{\sum\sum} -
b^{\dagger}_{\alpha} (\kb)
d^{\dagger}_{\alpha'}(\kpb)u^{\alpha \dagger}(\kb)v^{\alpha'}(\kpb)
\widehat{f}(-\omega_{\kb}-\omega_{\kpb})]   \}     .
\end{eqnarray}
Comparing with Eq.~\eqref{eq:Tf}, we conclude that
\begin{equation}
T_{f} = \frac{1}{2\pi}  \int_{-\infty}^{\infty}  d\mu\, \mu
\left(   \mathcal{O}^{\dagger}_\mu \mathcal{O}_\mu - S_{\mu}\II \right ),
\label{eq:T_f}
\end{equation}
which is
similar in form to the decomposition of the scalar field energy density employed 
in~\cite{FewsterEveson}. 
Next we compute the anticommutator
\begin{eqnarray}
\{ \mathcal{O}_{\mu i}^{\dagger} , \mathcal{O}_{\mu i}  \}  &=& \sum_{\alpha,
\alpha'} \sum_{\kb,\kpb}  \{ \widehat{g}(-\omega_{\kb}+\mu) \overline{ \widehat{g}(-\omega_{\kpb}+\mu)  
} \{ b_{\alpha}^{\dagger}(\kb), b_{\alpha'}(\kpb)   \} u^{\alpha \dagger}
(\kb) u^{\alpha'}(\kpb) \notag \\ 
&&\phantom{\sum_{\alpha,\alpha'} \sum_{\kb,\kpb}}
+ \widehat{g}(\omega_{\kb}+\mu) \overline{ \widehat{g}(\omega_{\kpb}+\mu)  
} \{ d_{\alpha}(\kb), d_{\alpha'}^{\dagger}(\kpb)   \} v^{\alpha \dagger}
(\kb) v^{\alpha'}(\kpb) \}
\end{eqnarray} 
where we have again implicitly summed over the spinor index $i$ and used the fact
that anticommutators
between the $b_{\alpha}(\kb)$ and $d_{\alpha}(\kb)$ vanish. Using
the anticommutation relations, the above expression reduces to 
\begin{eqnarray}
\{\mathcal{O}_{\mu i}^{\dagger} , \mathcal{O}_{\mu i} \} &=& \left(
\sum_{\alpha, \kb} |\widehat{g}(-\omega_\kb+\mu)|^{2}\, \|u^{\alpha}(\kb)\|^{2} +
\sum_{\alpha, \kb} |\widehat{g}(\omega_\kb+\mu)|^{2}\, \|v^{\alpha}(\kb)\|^{2}\right)\II
\notag \\
&=& ( S_{-\mu} + S_{\mu} )\II,
\label{eq:OdaggerO}
\end{eqnarray} 
where we have also used Eq.~\eqref{eq:normalisation} and the fact that
$|\widehat{g}(u)|$ is even (because $g$ is real). 
Splitting the $\mu$-integral of Eq.~\eqref{eq:T_f} into two pieces and
using Eq.~\eqref{eq:OdaggerO} gives
\begin{align}
T_{f} &= \frac{1}{2\pi} \int_{0}^{\infty} d\mu \, \mu (   \mathcal{O}_{\mu
i}^{\dagger}
  \mathcal{O}_{\mu i} - S_{\mu}\II   ) + \frac{1}{2\pi}
\int_{-\infty}^{0} d\mu \, \mu [ (S_{\mu} + S_{-\mu}   )\II  - 
\mathcal{O}_{\mu i} \mathcal{O}_{\mu i}^{\dagger} - S_{\mu}\II   
] \notag \\
&= \frac{1}{2\pi} \int_{0}^{\infty} d\mu \, \mu 
 (     \mathcal{O}_{\mu i}^{\dagger}  \mathcal{O}_{\mu i} - S_{\mu}\II  )
+ \frac{1}{2\pi} \int_{-\infty}^{0} d\mu \, \mu  (S_{-\mu}\II - \mathcal{O}_{\mu
i} \mathcal{O}_{\mu i}^{\dagger} ). 
\end{align}
Now, for any $\mu\geqslant 0$, we have $\mu \langle \mathcal{O}_{\mu i}^{\dagger}
\mathcal{O}_{\mu i} \rangle_{\psi}  \geqslant 0$, while for any
$\mu \leqslant 0$, we have $-\mu \langle \mathcal{O}_{\mu i} \mathcal{O}_{\mu
i}^{\dagger} \rangle_{\psi} \geqslant 0$
for all physically reasonable states $\psi$~\footnote{In a fully rigorous formulation (cf.~\cite{CJFVerch}),
one would expect the QWEI to hold for all Hadamard states.}.  Thus,
\begin{eqnarray}
{\langle T_{f} \rangle}_{\psi} &\geqslant&  -\frac{1}{2\pi}
\int_{0}^{\infty} d\mu \, \mu S_{\mu} + \frac{1}{2\pi} \int_{-\infty}^{0}
d\mu \, \mu S_{-\mu} \notag \\
&=& -\frac{1}{\pi}
\int_{0}^{\infty} d\mu \, \mu  S_{\mu}, 
\label{eq:bound}
\end{eqnarray}
so we have obtained a state-independent lower bound on $\langle T_{f}
\rangle_{\psi}$. We now proceed to calculate the right-hand side of Eq.~\eqref{eq:bound} and show
that it is finite.  Inserting the definition Eq.~\eqref{eq:Smu}
of $S_{\mu}$ into Eq.~\eqref{eq:bound}, we find
\begin{equation} 
\langle T_f \rangle_\psi \geqslant -\frac{2}{\pi}\int_{0}^{\infty}
d\mu\, \mu \frac{1}{V} \sum_\kb |\widehat{g}(\omega_{\kb} +\mu)|^2.
\end{equation}
Taking the continuum limit
\begin{equation} \frac{1}{V}\sum_{\kb} \longrightarrow \int \frac{d^{3}\kb}{(2\pi)^3},
\end{equation}
the inequality becomes
\begin{eqnarray}
 \langle T_f \rangle &\geqslant & -\frac{2}{\pi} \int_{0}^{\infty} d\mu\,
\mu \int_{0}^{\infty} \frac{d^{3}\kb}{{(2\pi)}^{3}}  |\widehat{g}(\omega_{\kb}+\mu)|^2 \notag \\
&=& -\frac{2}{\pi} \int_{0}^{\infty} d\mu\, \mu \frac{1}{8\pi^3} 4\pi
\int_{0}^{\infty} dk\, k^{2}      
|\widehat{g}(\omega_{k}+\mu) |^2 \notag \\
&=&  -\frac{1}{\pi^3} \int_{0}^{\infty} d\mu\, \mu \int_{m}^{\infty}  d\omega\,
\omega\sqrt{\omega^2 - m^2}   
|\widehat{g}(\omega+\mu) |^2,
\label{eq:QI}
\end{eqnarray}
where we have made a change of variables in the integration
to spherical polar coordinates, integrated over the angular
variables and changed variable again 
to the energy $\omega=\sqrt{k^2+m^2}$. 
The next step is to make a further change of variable
\begin{equation}
u=\omega + \mu, \ \ v=\omega, 
\end{equation}
after which the quantum inequality Eq.~\eqref{eq:QI} becomes
\begin{eqnarray}
 \langle T_{f} \rangle &\geqslant&  -\frac{1}{\pi^3}  \int_{m}^{\infty} du\, |\widehat{g}(u)|^{2}
 \int_{m}^{u} dv\, v(u-v) \sqrt{v^2 - m^2} \notag \\
&=& -\frac{1}{\pi^3} \int_{m}^{\infty} du\, |\widehat{g}(u)|^2
\left(  \frac{1}{3}  u  (u^2 - m^2)^{3/2} -
\frac{1}{4}u^{4}  Q_{3} \left( \frac{u}{m} \right) \right),
\label{eq:QImass} 
\end{eqnarray}
where the function $Q_{3}(x)$ is given in Eq.~\eqref{eq:Q3}.

Using the translational invariance of the theory to move from a
worldline $(t,\boldsymbol{0})$ to $(t,\boldsymbol{x_0})$,
we have therefore established the bound
\begin{equation}
\int  dt\, {\langle {:}T_{00}{:} \rangle}_{\psi} (t, \boldsymbol{x_0})  {g(t)}^{2}
\geqslant - \frac{1}{12{\pi}^{3} } \int_{m}^{\infty} du\, |\widehat{g} (u) |^{2} u^{4}
Q_{3}^{D}(u/m),
\label{eq:QIfinal}
\end{equation}
for any $\boldsymbol{x_0}$, where
\begin{equation}
Q_{3}^{D}(x) = 4   \left( 1 - \frac{1}{x^2} \right)^{3/2} - 3 Q_{3}(x),
\end{equation}
as claimed in the Introduction.
This bound is clearly finite for any smooth compactly supported $g$,
as $u^{4} Q_{3}^{D}(u/m)$ grows like $u^4$
for large $u$, while $|\widehat{g}(u)|^{2}$ decays faster
than any inverse polynomial in $u$ as $u\to\infty$.

\section{Massless Case}

\begin{figure}
\begin{center}
\rotatebox{270}{\scalebox{0.6}{
\includegraphics{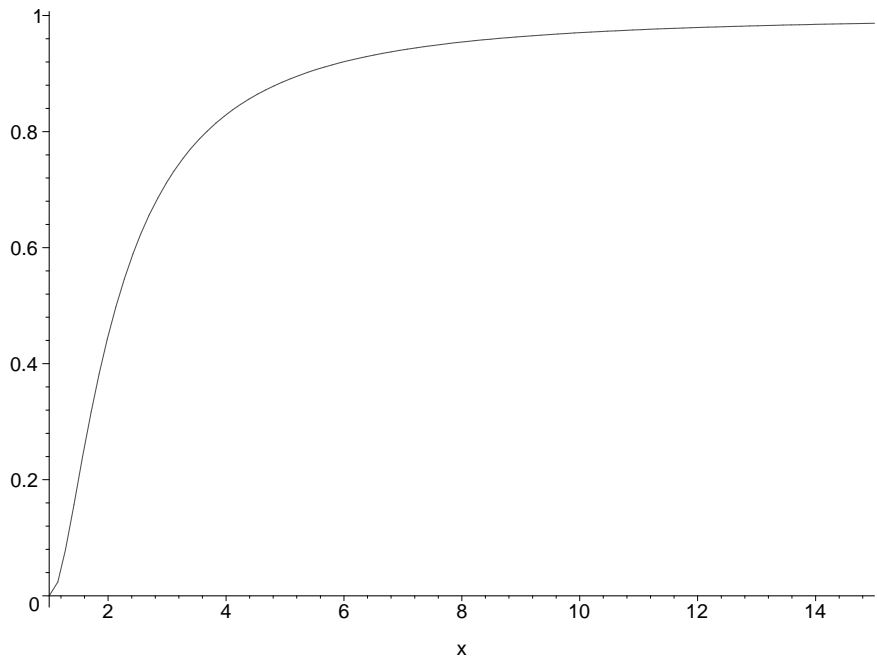} 
}}
\end{center}
\caption{\label{fig:Q3D} Plot of $Q_3^D(x)$.}
\end{figure}

We now look at the simplest case of the general QWEI
derived above; the massless case.  We then
compare our result for the Dirac field with those for the scalar field
derived by Ford and Roman~\cite{fordroman3,FordRoman97} and Fewster
and Eveson~\cite{FewsterEveson}. Substituting $m=0$ in the first line of
Eq.~\eqref{eq:QImass} and performing the integral over $v$, we
immediately find 
that the four-dimensional QWEI for massless fields takes the
form 
\begin{equation}
\langle T_f \rangle _\psi \geqslant
-\frac{1}{12\pi^3} \int_{0}^{\infty} du |\widehat{g}(u)|^2 u^4,
\end{equation}
which is also the $m\to 0$ limit of Eq.~\eqref{eq:QIfinal} because
$Q_{3}^{D}(x) \rightarrow 1$ as $x
\rightarrow \infty$. Since the integrand is an even function in $u$, we may extend the
range of integration to the whole of $\mathbb{R}$ to give
\begin{eqnarray}
\langle T_f \rangle _\psi &\geqslant& -\frac{1}{24\pi^3} \int_{-\infty}^{\infty} du
 |u^2\widehat{g}(u)| ^2  . \notag  \\
&=& -\frac{1}{12\pi^2}\int_{-\infty}^{\infty} dt \left| g''(t)
\right| ^{2},
\label{eq:masslessQWEI}
\end{eqnarray}
where we have used Parseval's theorem to yield a $t$-space
version of the quantum inequality in the last line. In passing, we
observe that, because $Q_3^D$ is bounded between $0$ and $1$ (see Fig.~\ref{fig:Q3D}), the
introduction of a mass only serves to tighten the QWEI bound.
Accordingly, Eq.~\eqref{eq:masslessQWEI} also holds for massive Dirac
fields. 

Comparison with the quantum inequality for the scalar and electromagnetic
fields in Minkowski space derived in~\cite{FewsterEveson,Pfenn}
shows that the massless Dirac bound is weaker than the corresponding 
scalar bound by a factor of $\frac{4}{3}$, and stronger than the
electromagnetic bound by a factor of $\frac{2}{3}$. Since none of these
bounds are optimal, it is difficult to draw definite conclusions from this
beyond the observation that the bounds are of comparable magnitude.

In particular, for the choice of a Lorentzian sampling function peaked
at $t=0$, with characteristic width $\tau$
\begin{equation}
L(t)= \frac{\tau}{\pi(t^{2} + \tau^{2})}
\end{equation}
in which $\tau$ sets the averaging timescale, it is straightforward to calculate
\begin{equation} 
\langle T_L \rangle _\psi \geqslant -\frac{1}{12\pi^2}
\int_{-\infty}^{\infty} dt    \frac{{(2t^2 - \tau^2)}^2\tau }{\pi{(t^2 +
\tau^2)}^5}
= -\frac{36}{2048\pi^2 \tau^4}, 
\end{equation}
(at least for those states for which the right-hand side is well-defined).
This is a slightly stronger result, by a factor of $\frac{3}{16}$, than
the quantum inequality derived by Ford and
Roman~\cite{fordroman3,FordRoman97} for the scalar field. We have
therefore explained why the specific classes of states considered
in~\cite{Vollick1,YuShu} satisfy the Ford--Roman QWEI. We also observe
that a traversable wormhole supported by a Dirac field would be subject to slightly
tighter constraints than those envisaged in~\cite{FR-worm}.

\section{Conclusion}

We have derived a new and explicit QWEI for the Dirac field of mass
$m\geqslant 0$ in four-dimensional Minkowski space, by straightforward
means. Our argument is related to the general (but inexplicit) QWEI
obtained in~\cite{CJFVerch},
in that Lemma~1 is at its heart; one may therefore regard it as
an improved and streamlined version of~\cite{CJFVerch} (for the particular
case of averaging along an inertial trajectory in Minkowski space) with
the added benefit of a closed-form expression for the bound. 
Although it is at present somewhat formal, we expect that our argument
can be made fully rigorous (by means of microlocal techniques,
cf.~\cite{AGWQI,CJFVerch}) and that it 
can shed further light on the general situation
in curved spacetimes, again leading to explicit bounds. 

Our argument is also related to that developed for the scalar field  
in~\cite{FewsterEveson,CJFTeo}, 
and which we briefly summarise in Appendix~\ref{appx:B}. In each case,
an identity for Fourier transforms (the convolution theorem or Lemma~1)
and the (anti)commutation relations are used to express the averaged
energy density as an integral of manifestly positive operators, modulo a
$c$-number term which eventually provides the lower bound. Extra care
is required in the Dirac case because the classical energy density is
unbounded both from above and below.

Finally, in contrast to Vollick's two-dimensional bound~\cite{Vollick2},
our QWEI is not expected to be the sharpest bound possible. Again, it is
hoped to address this issue elsewhere.

\appendix
\section{Proof of Lemma~1}

In this appendix, we wish to state and prove the following lemma
(originally proved in~\cite{CJFVerch} by a slightly different method):
If $f=g^2$ for real-valued $g\in C_0^{\infty}( \mathbb{R})$,
then
\begin{equation} 
(\lambda +\lambda')\widehat{f}(\lambda - \lambda') =  
\int_{-\infty}^\infty  \frac{d\mu}{\pi} \mu \widehat{g}(\lambda -\mu)
\overline{\widehat{g}(\lambda'-\mu)}. \label{eq:lem1res}
\end{equation}
To obtain this result, first notice that the left-hand side is the
momentum space integral kernel for the operator
\begin{equation}
A = p f+ fp
\end{equation}
where $f$ acts by multiplication and $p=-id/dx$. That is,
\begin{equation}
\widehat{A\psi}(\lambda) = 
 \int_{-\infty}^{\infty}\frac{d\lambda'}{2\pi}\, (\lambda + \lambda') \widehat{f} (\lambda-\lambda')
\widehat{\psi}(\lambda')
\label{eq:A1}
\end{equation}
holds (at least) for all smooth, compactly supported $\psi$. 
On the other hand, because $f=g^2$, we also have
\begin{equation}
A = pg^2+g^2p = 2gpg + [p,g]g+g[g,p] = 2gpg
\end{equation}
because $[g,p]=ig'$, another multiplication operator. Written as an
integral operator, we therefore have
\begin{equation}
\widehat{A\psi}(\lambda) = 
2\int_{-\infty}^\infty \frac{d\mu}{2\pi}\, \widehat{g}(\lambda-\mu)\mu 
\int_{-\infty}^{\infty}\frac{d\lambda'}{2\pi}\,\widehat{g} (\mu-\lambda')
\widehat{\psi}(\lambda').
\label{eq:A2}
\end{equation}
Since both~\eqref{eq:A1} and~\eqref{eq:A2} hold for all smooth compactly
supported $\psi$, the two integral kernels must also agree, therefore
yielding~\eqref{eq:lem1res}, using the fact that 
$\widehat{g} (\mu-\lambda')=\overline{\widehat{g}(\lambda'-\mu)}$
because $g$ is real.

\section{Sketch of the scalar field argument}
\label{appx:B}

For ease of comparison with our Dirac QWEI, we briefly summarise the
scalar field argument of~\cite{FewsterEveson,CJFTeo}. (Notation differs
slightly, and we have specialised to four-dimensional Minkowski space.) This argument
begins by observing that the averaged energy density may be written as a
sum of terms of the form
\begin{eqnarray}
S^\pm&=&\frac{1}{2}\int\frac{d^3\kb}{(2\pi)^3}\,\frac{d^3\kpb}{(2\pi)^3}\,\Big\{
\widehat{f}(\omega_\kpb-\omega_\kb)
\overline{p(\kb)} p(\kpb)
a^\dagger(\kb) a(\kpb) \nonumber\\
&&\indent\indent \pm \widehat{f}(\omega_\kb+\omega_\kpb)
p(\kpb)p(\kb)a(\kpb)a(\kb) + \hbox{h.c.}\Big\}\,,
\end{eqnarray}
where $a(\kb)$ and $a^\dagger(\kb)$ are the annihilation and creation
operators of the scalar field, and $p(k)$ is polynomially bounded in $k$.
Introducing operators 
\begin{equation}
{\cal O}^\pm(\omega)=\int\frac{d^{3}\kb}{(2\pi)^{3}}\Big\{
\overline{\widehat{g}(\omega-\omega_\kb)}p(\kb) a(\kb)\pm
\overline{\widehat{g}(\omega+\omega_\lambda)}\overline{p(\kb)} a^\dagger(\kb)\Big\},
\end{equation}
a calculation using the commutation relations satisfied by the
annihilation and creation operators and the convolution theorem in the
form
\begin{equation}
\int_{-\infty}^\infty \frac{d\omega'}{2\pi}
\widehat{g}(\omega-\omega')\widehat{g}(\omega') =
\widehat{g^2}(\omega)=\widehat{f}(\omega)
\end{equation}
(and the fact that $\widehat{g}(u)=\overline{\widehat{g}(-u)}$ as $g$ is real)
yields the expression
\begin{equation}
\int_{0}^{\infty}\frac{d\omega}{2\pi} \mathcal{O}_{\omega}^{\pm\dagger} 
\mathcal{O}_{\omega}^{\pm}
 = S^{\pm} + \int_{0}^{\infty} \frac{d\omega}{2\pi} \int \frac{d^{3}\kb}{{(2\pi)}^{3}} 
|\widehat{g}(\omega + \omega_\kb)|^{2} |p(\kb)|^{2}. 
\label{eq:prebound}
\end{equation} 
Since the right-hand side of Eq.~\eqref{eq:prebound} is (formally) a manifestly positive
operator, we conclude that the expectation value of $S^{\pm}$ in any
quantum state $\psi$ must satisfy the following bound:
\begin{equation}
\langle S^\pm\rangle_{\psi}   \geqslant
- \int_0^\infty \frac{d\omega}{2\pi} \int \frac{d^{3}\kb}{(2\pi)^{3}}  
|\widehat{g}(\omega+\omega_\kb)|^2 |p(\kb)|^2 
\end{equation}
for all (physically reasonable) states $\psi$. It follows that the
averaged energy density satisfies a state-independent lower bound. 
Further manipulation
leads to the bound of Eq.~\eqref{eq:scalarQI} with $\mathcal{S}=1$. 

\begin{acknowledgments}
The work of CJF was assisted by EPSRC
Grant GR/R25019/01 to the University of York.
\end{acknowledgments}
%\newpage


\begin{thebibliography}{99}

\bibitem{EGJ} H. Epstein, V. Glaser, and A. Jaffe, Nuovo Cimento {\bf 36}, 1016 (1965).

\bibitem{FR-worm} L.H. Ford and T.A. Roman, Phys. Rev. D {\bf 53}, 5496 (1996).

\bibitem{Ford78} L.H. Ford, Proc. Roy. Soc. Lond. {\bf A364}, 227 (1978).

\bibitem{fordroman3} L.H. Ford and T.A. Roman, Phys. Rev. D {\bf 51}, 4277 (1995).

\bibitem{FordRoman97} L.H. Ford and T.A. Roman, Phys. Rev.  D {\bf 55}, 2082 (1997).

\bibitem{flan} \'E.\'E. Flanagan, Phys. Rev. D {\bf 56}, 4922  (1997).

\bibitem{PfenningFord98} M.J. Pfenning and L.H. Ford, Phys. Rev. D {\bf 57}, 3489 (1998).

\bibitem{FewsterEveson} C.J. Fewster and S.P. Eveson, Phys. Rev. D {\bf 58}, 084010 (1998).

\bibitem{CJFTeo} C.J. Fewster and E. Teo, Phys. Rev D {\bf 59}, 104016 (1999).

\bibitem{AGWQI} C.J. Fewster, Class. Quantum Grav. {\bf 17}, 1897 (2000).

\bibitem{Pfenn} M.J. Pfenning, Phys. Rev. D {\bf 65}, 024009  (2002).

\bibitem{FewsterPfenning}
C.J. Fewster and M.J. Pfenning, 
`A Quantum Weak Energy Inequality for spin-one fields in curved spacetime',
{\tt arXiv:gr-qc/0303106}, to appear J. Math. Phys. (2003). 

\bibitem{Helfer2} A.D. Helfer, 
`The Hamiltonians of Linear Quantum Fields: II. Classically Positive
Hamiltonians', {\tt arXiv:hep-th/9908012}.

\bibitem{Flan2} \'E.\'E. Flanagan, Phys. Rev. D {\bf 66}, 104007 (2002). 


\bibitem{Vollick2} D.N. Vollick, Phys. Rev.  D {\bf 61},  084022  (2000).  

\bibitem{CJFVerch} C.J. Fewster and R. Verch, Commun. Math. Phys. {\bf 225}, 331 (2002).

\bibitem{Vollick1} D.N. Vollick, Phys. Rev. D {\bf 57}, 3484 (1998).

\bibitem{YuShu} H. Yu and W. Shu, Phys. Lett. B {\bf 570}, 123 (2003).

\bibitem{Belinfante} F. Belinfante, Physica (Amsterdam) {\bf 6}, 887 (1939). 





\end{thebibliography}
\end{document}